# TO RE-CONSIDER THE ONE-WAY SPEED OF LIGHT USING FIZEAU-TYPE-COUPLED-SLOTTED-DISKS

Md. Farid Ahmed[1,*], Brendan M. Quine[1,2], Stoyan Sargoytchev[3], and A. D. Stauffer[2]

[1]Department of Earth and Space Science and Space Engineering, York University, 4700 Keele Street, Toronto, Ontario, Canada-M3J 1P3.
[2]Department of Physics and Astronomy, York University, 4700 Keele Street, Toronto, Ontario, Canada-M3J 1P3.
[3]Department of Computer Science and Engineering, York University, 4700 Keele Street, Toronto, Ontario, Canada-M3J 1P3.

[*]Address of corresponding author:
Department of Earth and Space Science and Space Engineering,
York University,
4700 Keele Street, Toronto, Ontario, Canada-M3J 1P3
Phone: 647-449-8320
E-mail: mdfarid@yorku.ca

## Shortened version of the title

TO RE-CONSIDER THE FIZEAU-TYPE ONE-WAY ISOTROPY TEST

## Abstract

The isotropy of the speed of light – the fundamental postulate of Special Relativity (SR) constrains conceptions of time, space and the existence of a preferred cosmological reference frame. Consequently, this phenomenon has been subject to considerable experimental scrutiny. Most isotropy tests are two-way Michelson-Morley type tests which established the isotropy of the two-way speed in 1881. These approaches provide no experimental limit for the one-way (single-trip) isotropy of the speed of light which is still unresolved. Here we consider Fizeau-type experiments to test the isotropy of the one-way speed of light. Our theoretical and experimental design suggests that our approach is 2600 times more sensitive than that of previous Fizeau-type experiments and 2000 times more sensitive than Michelson-Morley type two-way tests. We present our experimental methodology as well as initial calibration results for our experimental apparatus.

## 1. INTRODUCTION

The identification of a preferred frame of reference or any deviation from Special Relativity (SR) could assist physicists in the formulation of a quantum theory of gravity [1] – the unification of all fundamental forces in nature. Physicists are now testing the fundamental postulate of SR – the isotropy of the speed of light "with everything from enormous particle accelerators, to tiny electromagnetic traps that can hold a single electron for months, to bobs of metal twisting on the ends of long fibers," [1]. The isotropy of the speed of light is being tested using enhanced Michelson-Morley type experiments on the Earth as well as on the space station [2]. Most of these experiments utilize a two-way methodology which can establish only the round-trip averaged speed of light over closed paths [2, 3]. However, these approaches provide no experimental limit for the one-way (single-trip) isotropy of the speed of light which is still unresolved [2,4,5].

Clock-synchronization is one of the important considerations for the direct test of one-way speed of light. According to general test theory [6], the one-way velocity of light depends on the synchronization parameter. However, Will [7] showed that experiments which test the isotropy in one-way or two-way (round-trip) experiments have observables that depend on test functions but not on the synchronization procedure. He noted that "the synchronization of clocks played no role in the interpretation of experiments provided that one is careful to express the results in terms of physically measurable quantities." Hence the synchronization is irrelevant for our one-way speed of light test since we express our results in terms of physically measurable quantities.

To test the idea of the constancy of the speed of light unambiguously, we need an experiment which should be beautiful in its simplicity and sensitive enough to measure the hypothetical violation of the constancy of the speed of light. A mechanically synchronized-spinning-coupled-slotted-disk was first used by Armand Hippolyte Fizeau in an early work to measure the speed of light in 1849 [8]. This concept was improved and used by Marinov in his spinning-coupled-slotted-disks experiment [9 – 11]. Following [2, 12] we would like to re-consider the present measurement with mechanically synchronized Fizeau-type-coupled-slotted-disks according to a method previously derived by Marinov. The recent discussion in Technology Review [13] gave us a further motivation for the present test.

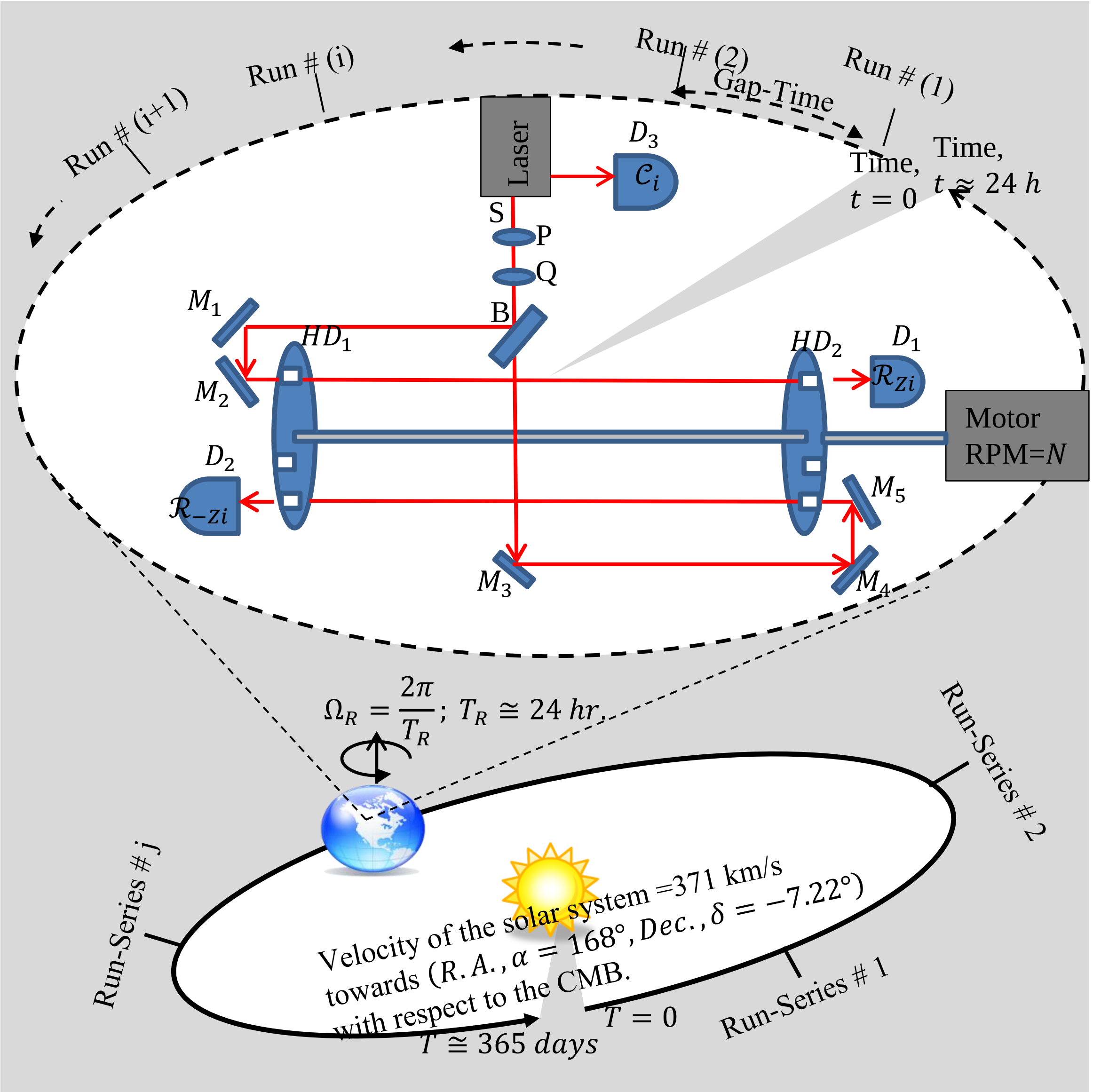

Fig. 1 Schematic of the basic setup of the experiment represents different Run # i and Run-Series # j. Legend: D ($D_1$, $D_2$, $D_3$) = Photodiodes; $\mathcal{R}_{zi}$, $\mathcal{R}_{-zi}$, $\mathcal{C}_i$= average responses of detectors $D_1$, $D_2$, $D_3$ respectively; M ($M_1$ .. $M_5$) = Mirrors; $HD_1$,$HD_2$ = Holed Disks; B= Beam Splitter; Q = Quarter Wave Plate; P = Polarizer; S = Laser Source. A Run-Series # (j) spans at least a 24-hour period with different Runs # (i). During each Run we turn on our motor with the speed of 3200 RPM and collect data for about 2 minutes.

As shown in Fig. 1, the experiment consists of a shaft coupled with two holed disks $(HD_1, HD_2)$ at its extremities which is driven by an electromotor. Light from a laser source (S) is divided into two beams by a beam-splitter (B) and adjustable mirrors $(M_1\ to\ M_5)$ direct the two beams to the opposite ends of the rotating shaft, so that the light rays can pass through the holes of the disks in mutually opposite directions and illuminate photo detectors $(D_1, D_2)$ placed at either end of the experiment. The responses of the photo detectors are collected for a couple of minutes at a motor speed $N$ RPM. The average of these collected responses from a detector $(D_1)$ is called the response $\mathcal{R}_{zi}$ for the light propagation along $z$-axis for any Run # $i$ in any Run-Series # $j$. Similarly, the average of the collected responses from a detector $(D_2)$ is called the response $\mathcal{R}_{-zi}$ for the light propagation along $-z$-axis. For our improved setup, we collect control response $\mathcal{C}_i$ which is the average of the collected responses from a detector $(D_3)$ at the same time of $\mathcal{R}_{zi}$ and $\mathcal{R}_{-zi}$. Also we use a polarizer (P) to reduce the optical feedback and a Quarter Wave Plate (Q) to convert linear polarization into circular polarization. There is a series of Runs # (1 .. $i$, $i$+1, ..) during the course of the Run-Series # $j$ for a 24-hour period during which the Earth makes a complete rotation. Following [2], we can derive the time dependent component of the velocity $v(t)$ of the laboratory relative to a reference frame along the direction of the light propagation. Therefore, according to a Galilean transformation, if the speed of light depends on the speed of the source or the speed of the observer then the differential responses $(R_{zi} - R_{-zi})$ should be proportional to the variation of $v(t)$ during the course of a Run-Series # $j$ for a 24-hour period. There will be different Run-Series # (1, .. $j$, ..) throughout the year in different seasons.

While Marinov claimed the experimental setup was "childishly simple and cheap", other researchers found it challenging [11]. We would like to support the claim that it is challenging due to following reasons:

(1) Laser Stability: Following [14] the frequency instability $(S_\nu^{-1})$ of a typical gas laser is the sum of the fractional change in cavity length $\left(\frac{\Delta L}{L}\right)$ and the fractional change in refractive index $\left(\frac{\Delta n}{n}\right)$. Therefore, the instability can be written as

$$S_v^{-1}(\tau) = \frac{\Delta L(\tau)}{L} + \frac{\Delta n(\tau)}{n} \quad (1)$$

where $\tau$ is the period of observation. For a change in temperature $\Delta T$, we can write

$$\frac{\Delta L}{L} = \alpha \Delta T \quad (2)$$

where $\alpha$ is the coefficient of thermal expansion. Therefore, the temperature must be stable to $(10^{-2})$°C to obtain a stability of $10^{8}$ for a typical gas laser [14]. Also frequency variations due to fluctuations in refractive index of the inverted population and electron density are important. Vibrations or sagging can represent another kind of instability [15]. For an example, an angular variation of a typical gas laser tube of as little as $\pm 5 \times 10^{-6}$ rad would cause a frequency shift $\frac{\Delta v}{v} = \pm 2.0 \times 10^{-8}$, which emphasizes the importance of a rigid and vibration free laser mount [14].

(2) External Effects: Changes in atmospheric pressure, humidity and ambient temperature which change the refractive index and length in the external optical path contribute to the long term drifts. Temperature control is also an important concern for the isotropy of light tests [16]. The observer's own body-heat or infrared radiation can produce an effect of the test results [17].

(3) Scattering Effects: The elastic scattering effects due to the different parts of the disks can cause significant disturbances in a measurement as described in Fig.2. In order to understand these effects, let us consider that a laser beam enters through a slit of a disk. Assuming the thickness of the chopping edge is responsible for the scattering, we can model the approximate radiating volume = $V(d_2, d_{th})$. Therefore, following [18] the contributions due to the scattering is $= V(d_2, d_{th}) \times \mathcal{K}(\vec{k}_L - \vec{k}_S)$, where $\mathcal{K}(\vec{k}_L - \vec{k}_S)$ is the Fourier component of the susceptibility corresponding to the scattering vector $(\vec{k}_L - \vec{k}_S)$, $d_2$ is the width of the laser beam and $d_{th}$ is the thickness of the disk.

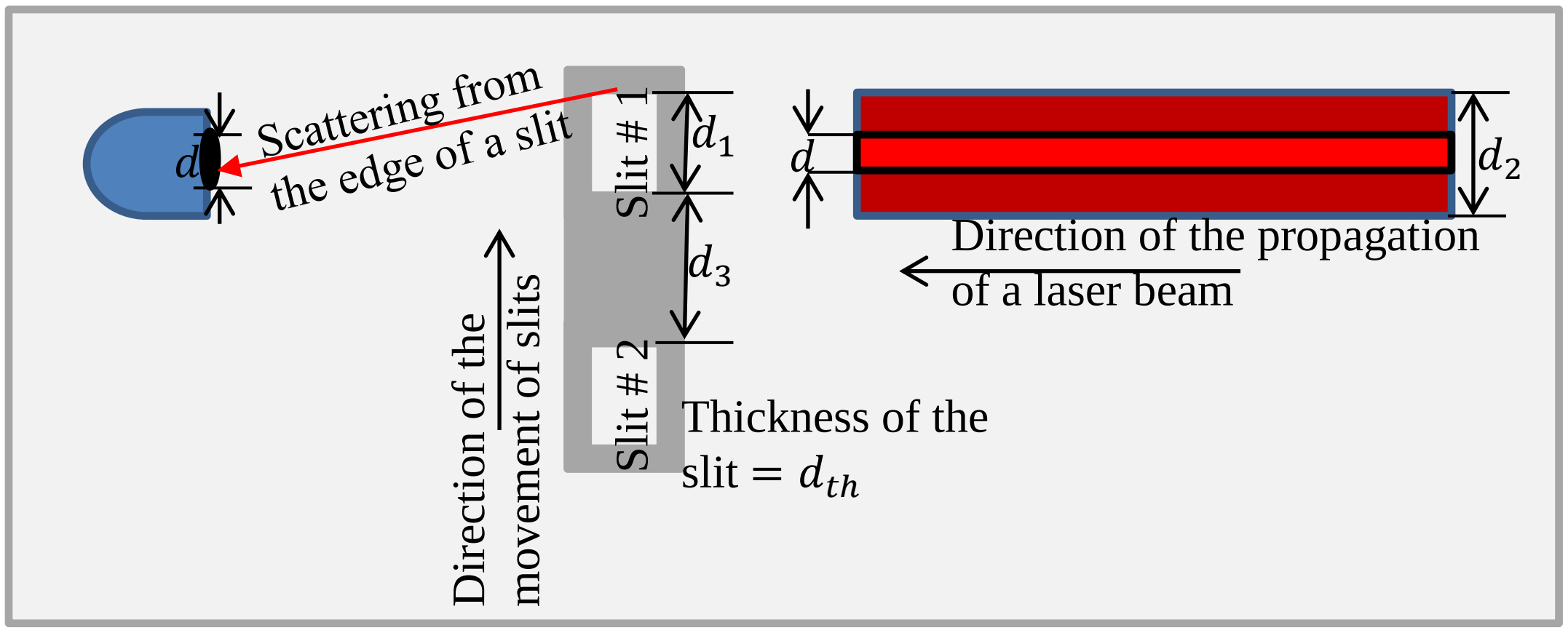

Fig.2 Schematic of the experimental configuration illustrates the investigation of light scattering by the slits.

The distance between two adjacent slits of the disk is $d_3 (> d_2)$. The width of the aperture of a detector is $d (< d_2)$ which is assumed as the width of a laser beam accepted by the detector in our interpretation in the next section. Ideally, we can assume that the parameters of the slits (Slit # 1, Slit # 2 and so on), of the two opposite laser beams and two detectors are exactly identical.

In order to minimize uncertainties due to the above challenges and also to make corrections, we propose some improvements to the basic experiment as described in the following sections.

## 2. A THEORETICAL INTERPRETATION

In this section we define theoretically the meaning of a response received by a detector. This response may be determined by amount of the laser beam that is incident on the detector per unit time. A visualization of the pattern received by the detectors after chopping by the disks is shown in Fig. 3 and Fig. 4.

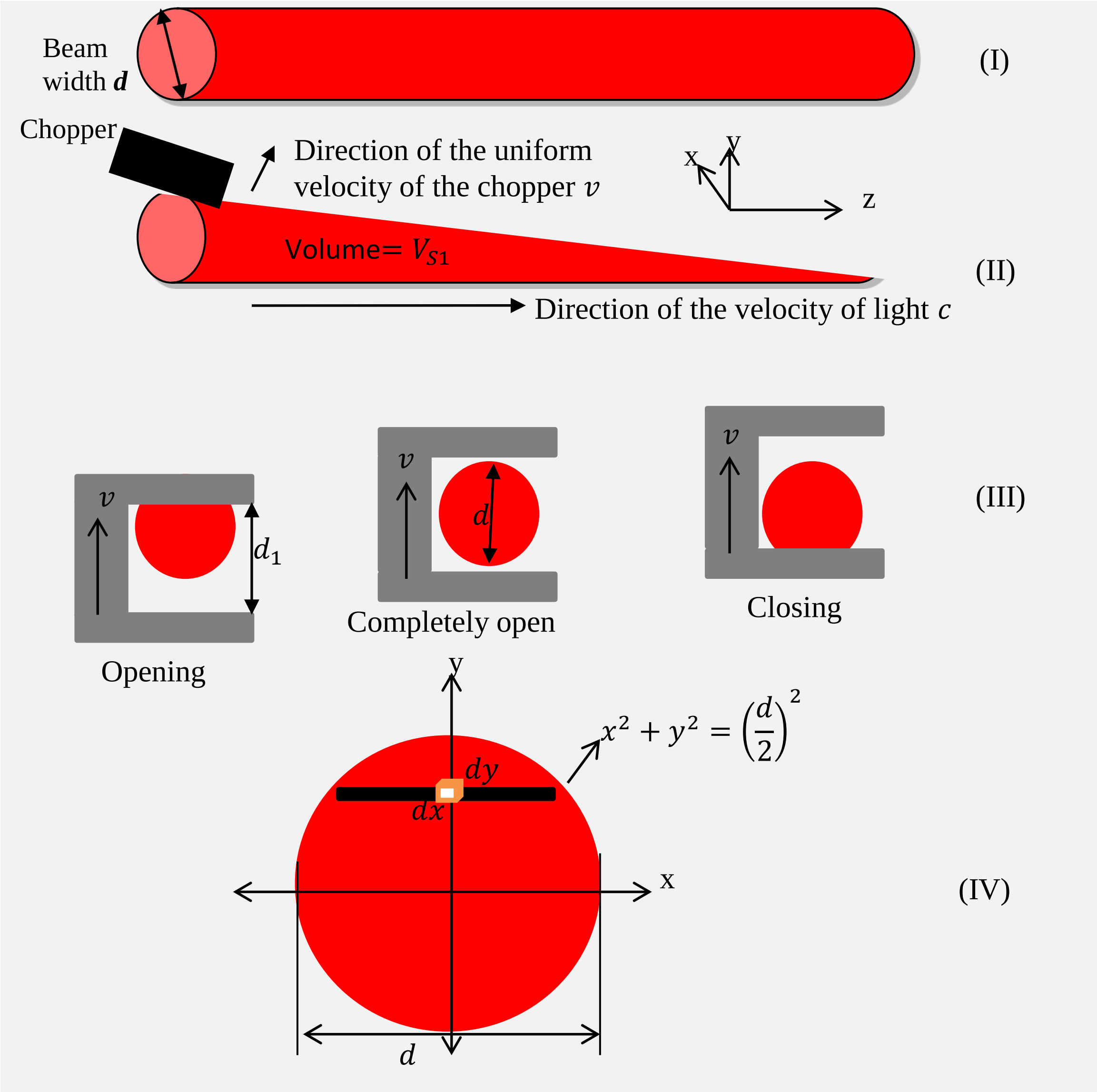

Fig 3 (I) Shape of a homogeneous laser beam of width $d$ before chopping in a frame $(x, y, z)$, (II) Shape of a homogeneous laser beam during chopping by a chopper of uniform velocity $v$, (III) Movement of a slit of width $d_1$ in the $xy$ plane, (IV) A projection of the laser beam on the $xy$ plane being chopped. The diameter of the beam $d$ depends on the shape and size of the aperture of the detector. The aperture is circular in this figure.

As shown in Fig. 3, we consider the laser beam that hits the photo detectors as homogeneous. As the laser illuminates the photo detectors, it passes through a circular aperture with diameter $d$ as shown in Fig. 5. Thus we can consider the shape of the beam to be a circular cylinder of diameter $d$. Let us assume that the direction of the velocity of light $c$ is along the $z$-axis and the direction of the uniform velocity of the chopper $v$ is along the $y$-axis. The beam of the laser is chopped twice by a slit of a disk for the same time duration $\frac{d}{v}$: firstly during opening, and secondly during closing. Also, assuming the width of a slit is $d_1 > d$, the whole beam will go through the slit for a duration of time $\frac{d_1 - d}{v}$. The distance between the two discs $L$ determines the amount of the beam that is blocked by the second disk.

In order to derive a general form of the volume $V_{S1}$, we consider a simple case using triple integration where the limits can be determined by drawing a projection of the laser beam in the $xy$-plane as shown in Fig. 3.

$$V_{S1} = \int_{-\frac{d}{2}}^{\frac{d}{2}} dy \int_{-\sqrt{\left(\frac{d}{2}\right)^2 - y^2}}^{\sqrt{\left(\frac{d}{2}\right)^2 - y^2}} dx \int_{\frac{y}{v}}^{\frac{d}{v}} c dt = K_{S1} c \qquad (3)$$

where, we have written $dz = cdt$, $c$ is the speed of light and $K_{S1}$ depends on the experimental parameters which are constant for a specific setup.

As shown in Fig. 4, the volume ($V_S$) of the segment (*S*) of a beam chopped by the first disk is equal to the sum of the volume of the segment (*S1*) of the beam from the time it starts entering the slit until it completely passes through the slit ($V_{S1}$), the volume of the segment (*S2*) of the whole beam passing through the slit ($V_{S2}$), and the segment (*S3*) of the beam from the time it starts to be blocked to the time it is completely blocked ($V_{S3}$):

$$V_S = V_{S1} + V_{S2} + V_{S3} \qquad (4)$$

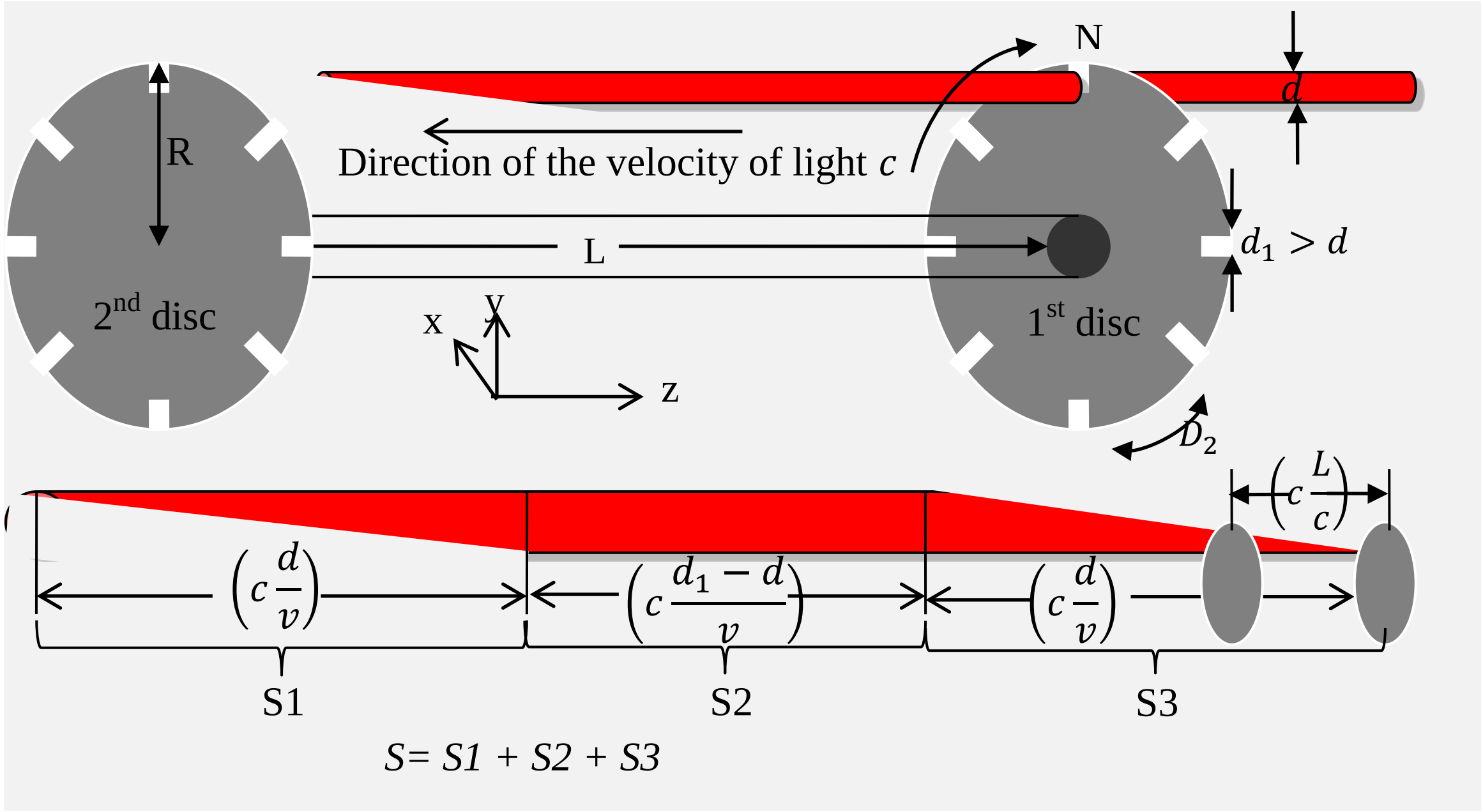

Fig 4: Illustration of a laser beam during chopping in the present experiment. The segment (*S*) of the chopped beam can be divided into three portions: *S1* = the segment of the beam from the time it starts to enter the slit until the time it completely passes, *S2* = the segment when whole beam passes through the slit, *S3* = the segment of the beam from the time it starts to be blocked to the time it is completely blocked. Legend: *L*= Distance between two disks; *R* = Distance of the chopping point from centre of the disc; *N* = Disk rotation speed per second; $d_1$ = width of the slit; *d* = aperture of the detector thorough which laser beam is entering; $D_2$ = distance between two slits.

For present experiment, the parameters are: $d = (2.0 \pm 0.05)$ mm; $d_1 = (5 \pm 0.05)$ mm; $R = (80 \pm 0.05)$ mm as shown in Fig. 5. These are necessary to derive the response $\mathcal{R}_{zi}$ of a detector. As $d$ is significant amount smaller than the distance between the centre of the shaft and the centre of the chopping point, we can assume a uniform velocity along the $y$-axis of the chopper at the chopping point. Therefore, as shown in Fig. 4, $V_{S1}$ is the half of the volume of a circular cylinder with diameter $d$ and length $\left(c\frac{d}{v}\right)$ which is equal to $V_{S3}$ by symmetry. And, $V_{S2}$ is the volume of a circular cylinder with diameter $d$ and length $\left(c\frac{d_1-d}{v}\right)$.

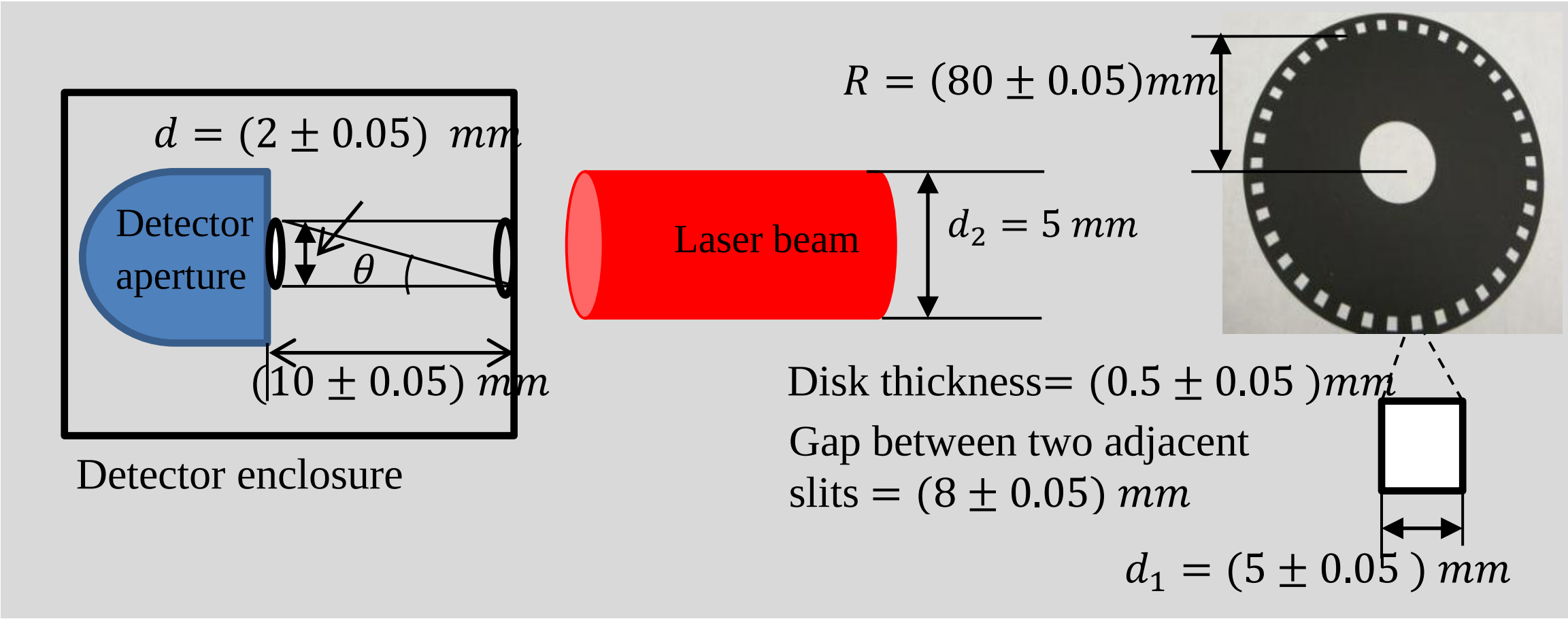

Fig. 5: Photograph of the disk (color: black to minimize reflection and scattering) with description which is being used in the present experiment. This disk was manufactured using laser technology. Different measureable physical parameters used in our experimental interpretation are shown.

Now, equation (3) can be derived for our present setup as:

$$V_{S1} = V_{S3} = \frac{1}{2}\pi\left(\frac{d}{2}\right)^2\left(c\frac{d}{v}\right) = K_{S1}c \tag{5}$$

Also, $V_{S2}$ can be derived as:

$$V_{S2} = \pi\left(\frac{d}{2}\right)^2\left(c\frac{d_1 - d}{v}\right) = K_{S2}c \tag{6}$$

Equation (4) can be derived using equations (5) and (6) as:

$$V_S = 2K_1c + K_2c = K_Sc \tag{7}$$

But a fraction of this volume $V_S$ is blocked by the second chopper. If one aligns the choppers such that the holes are lined up, the total volume $V$ of the laser beam being received by the detector can be calculated by subtracting the part of the segment (S4) of volume $V_{S4}$ blocked by the second disk.

As we have shown already in Fig. 4 that the form of the length of the blocked volume $V_{S4}$ by the second disk is $\left(c\frac{L}{c}\right)$, so we can derive volume of the blocked laser beam as:

$$V_{S4} \cong \pi\left(\frac{d}{2}\right)^2\left(c\frac{L}{c}\right) = K_{S4} \tag{8}$$

Using equations (7) and (8) the total volume $V$ of the laser beam being received by a detector can be derived as:

$$V = V_S - V_{S4} = K_S c - K_{S4} \tag{9}$$

where, $c$ is the speed of light and $K_{S1}$, $K_{S2}$, $K_S$ and $K_{S4}$ depend on the experimental parameters and which are constant for any specific setup. As we are interested in first order effects of the speed of light, equation (9) is sufficient for the present purpose.

The number of photons in the laser beam that hit the detector is proportional to the total volume $V$ of the chopped laser beam. If there are $n$ slits in the rotating disk and the rotational rate of the disk is $N$ cycles/sec then the total number of photons that hit the detector per sec will be proportional to $(nNV)$. From the above discussion we can approximate the response $\mathcal{R}$ of a detector for a laser beam propagating along the $z$-axis for "Run # $i$" as:

$$\mathcal{R} = K_1 V = K_2 c - K_3 \tag{10}$$

where $K_1$ is a proportionality constant which depends on $n$ and $N$; $K_2$ is a constant which depends on the parameters $n$, $N, d_1, d$ and $v$ of an experiment; $K_3$ is a constant which depends on the parameters $n, N, d$ and $L$.

If there is any variation in the speed of light then according to a Galilean transformation, the $c$ in equation (10) should be represented by $(c_0 + v(t))$ for one direction and $(c_0 - v(t))$ for the opposite direction where $v(t)$ is the component of the time dependent velocity of the laboratory along the direction of the laser beam relative to the absolute frame and $c_0$ is the speed of light relative to the absolute frame.

Therefore, using equation (10) we are able to derive the response ($\mathcal{R}_{zi}$) of the detector to the laser beam travelling in the $z$ direction and the response ($\mathcal{R}_{-zi}$) for the beam travelling in the opposite direction where $i$ is the "Run # $i$" in any Run-Series # $j$ for a 24-hour period as shown in Fig. 1. The responses ($\mathcal{R}_{zi}$), ($\mathcal{R}_{-zi}$) and the differential response ($\mathcal{R}_{zi} - \mathcal{R}_{-zi}$) can be derived as follows:

$$\mathcal{R}_{zi} = K_z[c_0 + v(t)] - K_z' \tag{11}$$

$$\mathcal{R}_{-zi} = K_{-z}[c_0 - v(t)] - K_{-z}' \tag{12}$$

$$\mathcal{R}_{zi} - \mathcal{R}_{-zi} = K_z[c_0 + v(t)] - K_{-z}[c_0 - v(t)] - K_z' + K_{-z}' \tag{13}$$

where the constants $K_z$, $K_{-z}$, $K_z'$ and $K_{-z}'$ denote constants for the two detectors. Ideally we would take $K_z \cong K_{-z}$ and $K_z' \cong K_{-z}'$. However, it is challenging to make two identical detectors with exactly the same apertures and responses. In order to avoid these challenges, we normalize the responses derived in equations (11 – 13).

Before we perform a derivation for normalization in general, we discuss $v(t)$. We derived the time dependent components of the velocity $v(t)$ of the laboratory along the direction of the light propagation in [2] assuming the Cosmic Microwave Background (CMB) is the rest frame of the universe. This derivation can help us to understand the shape of the change of velocity of the laboratory relative to the rest frame. For example, following the propagation direction of light in our laboratory in the North-South, we derive the time dependent component of the velocity of the laboratory relative to the rest frame along the direction of light propagation as follows:

$$\begin{aligned} v(t) = &\{\cos(\chi)\cos(\Omega_S t)\}\{-V_O \sin(\Omega_O t) + V_S \cos(\alpha)\cos(\delta)\} \\ &+ \{\cos(\chi)\sin(\Omega_S t)\}\{V_O \cos(\varepsilon)\cos(\Omega_O t) + V_S \sin(\alpha)\cos(\delta)\} \\ &+ \{-\sin(\chi)\}\{V_O \sin(\varepsilon)\cos(\Omega_O t) - V_S \sin(\delta)\} \end{aligned} \tag{14}$$

where $\chi =$ co-latitude, $\alpha =$Right Ascension, $\delta =$Declination, $\varepsilon =$the angle between the ecliptic and the Sun centered Celestial Equatorial plane, $\Omega_S =$ sidereal angular rotational frequency $\left(= \frac{2\pi}{23\,h\;56\,min} \cong 4.18 \times 10^{-3}\; deg.s^{-1}\right)$, $V_R =$ the velocity due to the Earth's rotation about its axis depending on the geographical latitude $(0 \le V_R \le 4.5 \times 10^2 ms^{-1})$, $\Omega_O =$the Earth is orbiting relative to the Sun with the angular frequency $\left(= \frac{2\pi}{1\,yr} \cong 1.14 \times 10^{-5}\; deg.s^{-1}\right)$, $V_O =$the velocity due to the Earth's orbital motion relative to the Sun $(\approx 3 \times 10^4\; ms^{-1})$ and $V_S =$the velocity of the solar system towards $(\alpha, \delta)$ relative to the rest frame. For the CMB as the rest frame we take $V_S =$the velocity of the solar system towards $[(\alpha, \delta) = (168°, -7.22°)]$ relative to the CMB $(\approx 3.71 \times 10^5 ms^{-1})$.

Based on above discussion of $v(t)$ in equation (14) [where we neglect time dependence of $\Omega_O$] and using [2, 19], we can derive the general form for $v(t)$ as follows:

$$v(t) = v_0 + v_1(t) \tag{15}$$

where $v_0$ is the constant term which can present along earth's rotation axis and $v_1(t)$ is the sinusoidal term which is perpendicular to the axis.

As shown in Fig. 1, a Run-series consists of a series of Runs # *(*1 .. $i$, $i$+1, *..)* and it lasts for a 24-hour period (a day, $T$) when the Earth makes a complete rotation. Therefore, considering an approximation where the change in $\Omega_O$ is negligible in equation (14) we can derive an average response $\bar{\mathcal{R}}_z$ for any Run-Series *#j* as follows:

$$\bar{\mathcal{R}}_z = \frac{1}{T}\int_0^T \{K_z[c_0 + v_0 + v_1(t)] - K'_z\}\, dt = K_z(c_0 + v_0) - K'_z \tag{16}$$

where the integration in equation (16) is an approximation of the average obtained by summing over on measurements for a period of 24-hours. Similarly, we can derive an average response for opposite direction: $\bar{\mathcal{R}}_{-z} = K_{-z}(c_0 - v_0) - K'_{-z}$.

Using equations (11), (15) and (16) we can derive the normalized responses $\mathcal{R}_{zi_N}$ for the response $\mathcal{R}_{zi}$ as follows:

$$\mathcal{R}_{zi_N} = \frac{\mathcal{R}_{zi}}{\bar{\mathcal{R}}_z} = \frac{K_z[c_0 + v_0 + v_1(t)] - K_z'}{K_z[c_0 + v_0] - K_z'} \approx 1 + \frac{v_1(t)}{c_0} + \mathcal{O}\left(\frac{1}{c_0^2}\right) \tag{17}$$

Similarly, using (12) the normalized response $\mathcal{R}_{-zi_N}$ for the opposite direction is:

$$\mathcal{R}_{-zi_N} = \frac{\mathcal{R}_{-zi}}{\bar{\mathcal{R}}_{-z}} = \frac{K_{-z}[c_0 - v_0 - v_1(t)] - K'_{-z}}{K_{-z}[c_0 - v_0] - K'_{-z}} \approx 1 - \frac{v_1(t)}{c_0} + \mathcal{O}\left(\frac{1}{c_0^2}\right) \tag{18}$$

As we are interested in the first order effects, therefore, after omitting 2nd and higher order and using equations (17) and (18), we can derive the normalized differential responses $\mathcal{R}_{diff_i_N}$ as follows:

$$\mathcal{R}_{diff_i_N} = 2\frac{v_1(t)}{c_0} \tag{19}$$

These first order effects derived in equations (17), (18) and (19) are being tested with the experiment described in the present paper. It is located in the Space Engineering Laboratory at the Centre for Research in Earth & Space Science (CRESS), York University, Toronto, Canada.

## 3. EXPERIMENTAL SETUP

We have improved the experimental setup over the basic version shown in Fig. 1. This is presented in Fig. 6a and 6b.

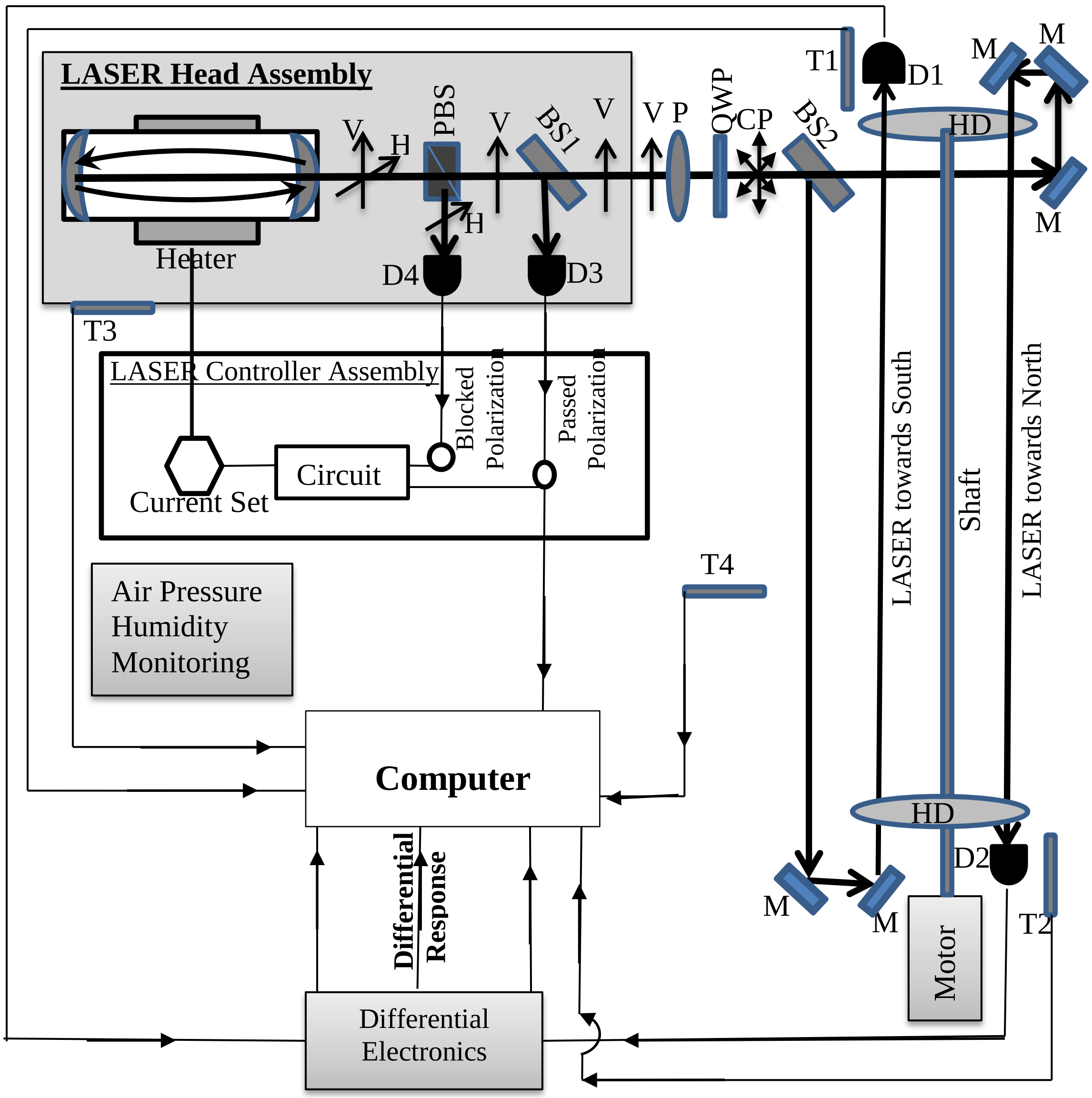

Fig. 6a Block Diagram of the improved experimental setup including NL-1 Frequency Stabilized He-Ne Laser, (Newport). Legend: D (D1, D2, D3, D4) = Photodiodes; T (T1 to T4) = Temperature sensors; M = Mirrors; HD = Holed Disks; BS(BS1, BS2)= Beam Splitters; QWP = Quarter Wave Plate; PBS = Polarizing cube Beam Splitter; V (Vertical), H (Horizontal) = Polarization mode; P=Polarizer; CP=Circular Polarization.

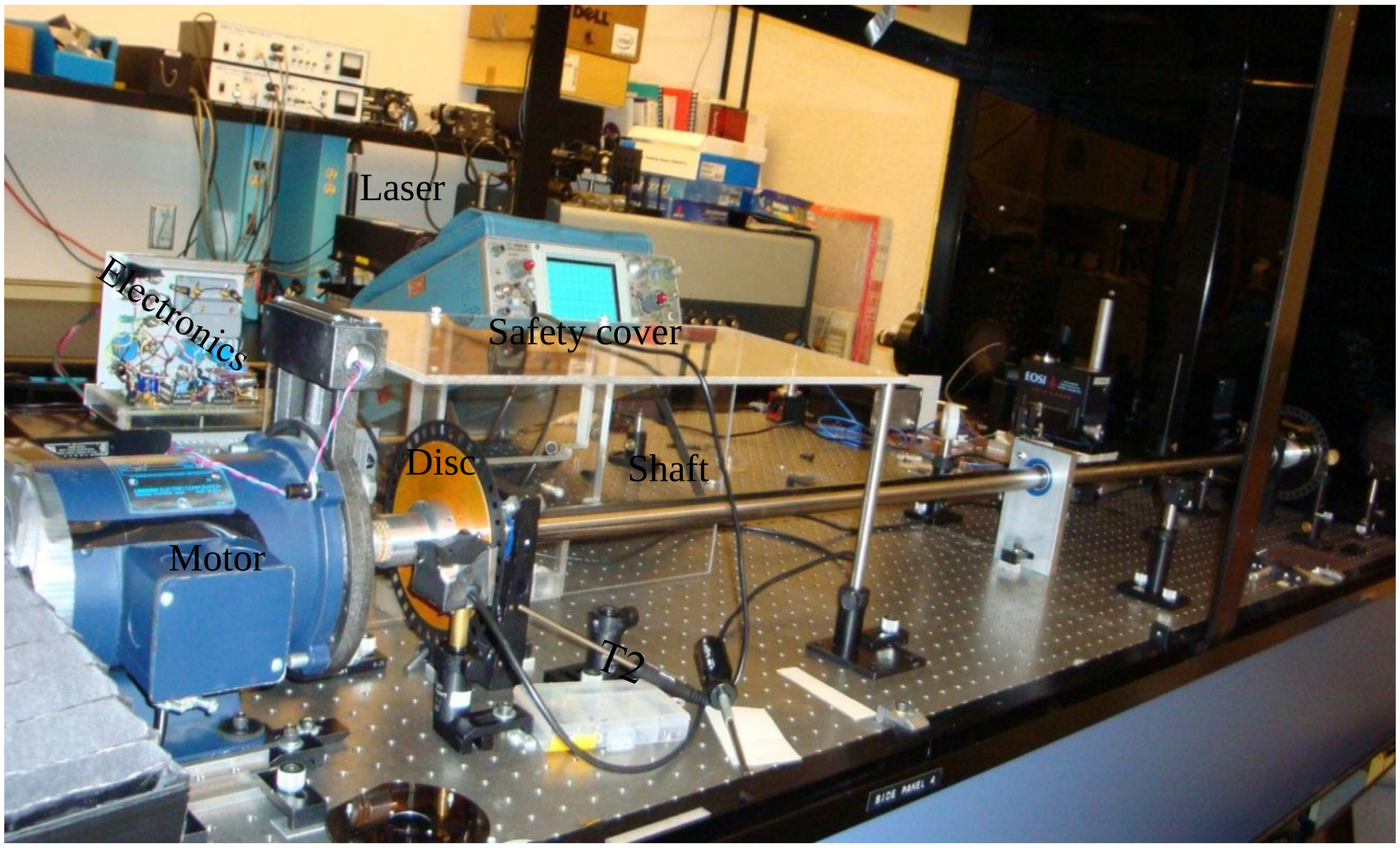

Fig. 6b Photograph of the total setup at the Centre for Research in Earth & Space Science (CRESS), York University, Toronto, Canada. The setup is on an optical table (KNS Series Table Top, Newport Corporation).

An NL-1 ultra-stable [20] frequency stabilized He-Ne Laser is used in these measurements. Two orthogonally-polarized modes (H and V) oscillate simultaneously in the laser gain tube. The laser beam passes through a polarizing cube beam splitter (PBS), which diverts one mode (H) to illuminate a photodiode (D4). The polarized beam (V) passes through a non-polarizing dielectric beam splitter (BS1), which diverts about 20% of the beam to another photodiode (D3). The outputs of the two photodiodes (D3 and D4) – representing the intensities of the two polarizations modes in the raw beam – are monitored by a differential comparator circuit which drives a thermal servo mechanism that keeps the laser cavity tuned to a particular resonance frequency [20]. These He-Ne lasers were studied in the laboratory at the Joint Institute for Laboratory Astrophysics (JILA). They reported the stability was better than 1 part in $10^{10}$ over the periods of about 1 hour and 1 part in $10^{8}$ over 1 year [21]. Also the Allan variance [22] plot for a typical red side lock for He-Ne laser was presented in the same report in [21]. However, environmental conditions must be stabilized in order to get a stable laser. The experiment operates for an extended period of at least 24 hours for one run data set. We monitor temperatures (T1, T2, T3, T4), humidity and atmospheric pressure in the laboratory. Also, in our improved setup we collect the output of the photodiode (D3) (Passed Response (Control)) which records the output intensity of the laser. Changes in internal laser intensity due to external variations of environmental conditions can be identified using this control signal.

Both long- and short-term stability can be maintained more easily by reducing vibration effects on critical components. Therefore, the Laser is mounted on a vibration-isolated table to reduce mechanical drift. Also, short-term stability is highly dependent on the amount of optical feedback [23], which is reduced by adjusting the polarizer (P). Also, in order to convert linear polarization (V) into Circular Polarization (CP) a Quarter Wave Plate (QWP) is used. This QWP requires proper orientation to get maximum circularly polarized light. We rotate the polarizer using a motor for any orientation of the QWP and observe the signal using a Power meter. In the optimized orientation of the QWP, the signal response in the power meter is almost constant. The orientation of the QWP is fixed for the rest of the experiment. Circular polarization of the laser reduces intensity variations due to mirror reflections in the optical path.

The laser beam is divided into two beams by a beam-splitter (BS2) and adjustable mirrors (M) direct the beams to opposite ends of the rotating shaft, so that the beams pass through the holes of the discs in mutually opposite directions and illuminate photodiodes (D1, D2). In order to test alignment, we first observe both signals of the photo detectors as shown in the Fig. 7. After observing and aligning both signals using an oscilloscope, we turn on our differential electronics that can converts photocurrent into potential difference and calibrate the zero differential response level. The conversion factor from photocurrent into voltage for this electronic device is $0.15\ \mu\mathrm{A}$ into $1\ \mathrm{V}$. Our setup records differentials of $0.01\ \mathrm{mV}$. Consequently, current changes as small as $\approx 0.0015\ \mathrm{nA}$ can be measured. By comparison, Marinov [11] claims that the “most sensible scale unit of the Austrian Norma galvanometer (that he was using), is 10 nA”. The sensitivity of the present Fizeau type experiment is linear with rotation speed, disk radius, disk separation and number of slits. Consequently, compared with Marinov, we compute our mechanical sensitivity to speed of light changes as $\left(\frac{3200}{12000} \times \frac{0.08}{0.12} \times \frac{1.755}{1.2} \cong 0.26\right)$ neglecting the effect of the number of slits which is unknown for the Marinov experiment. Therefore, we estimate that over all our setup is 2600 times more sensitive than this previous study [11]. A comparison of these different parameters of the present and previous experiments is presented in Table 1.

Table 1: Comparison of different parameters.

| **Specifications** | **Marinov [11] experiment** | **Present experiment** |
|---|---|---|
| The distance between the holed disks, $L$ | 1.2 m | 1.755 m |
| The distance between the center of the shaft to the center of the slits, $R$ | 0.12 m | 0.08 m |
| Number of slits, $n$ | (unknown) | 38 |
| Driving motor speed, $N$ | 12000 RPM | 3200 RPM |
| Photo current sensitivity | 10 nA | 0.0015 nA |

According to our approach, we adjust the alignment to synchronize the detector responses at slow rotation speeds as shown in Fig. 7. The previous studies by Marinov [9 – 11] followed a different approach: the alignment of the disks was such that when one disc closes then other opens. However, the interpretation of this experimental approach is unclear to us and unreliable data in our experiment was observed based on this approach.

A typical measurement is shown in Fig. 8. The deviation after about 2 minutes is due to temperature difference caused by the motor as shown in Fig. 9. The disturbances arise in part from thermally generated density fluctuations in the medium at constant pressure. It affects the two detectors unequally since the motor is situated at one end of the apparatus. Also, due to the interaction between a light waves and a sound waves traveling in the transmission medium create planes of higher and lower air density, and can cause scattering of light [24]. We used a special setup described in [25] to avoid the air flow fluctuations due to the motor. Following our approach, the typical resulting differential signal shown in Fig. 8 is reliable for $\leq 2$ miniutes. The edges of the slits/holes of the rotating disks can cause scattering effects as described in Fig. 2. Therefore, we set our detectors in enclosures with shielding as shown in Fig. 5. The paths of the laser beams received by the detectors are connected by the black plastic cylindrical tubes. These tubes have internal diameter which are equal to the aperture of the detectors and the lengths are $\approx 10$ mm which help us to reduce the scattering disturbances. Considering the scattering angle ($\theta \approx 11°$) in our setup as shown in Fig. 5, we estimate 4% of the total scattering which can cause disturbances based on solid angle produced. However, we tested the scattering effects by changing the width ($d_2$) of a laser beam for different discs (also, changing sizes, colors) and did not see any significant change. Therefore, we conclude that the present setup – using the black disks with slits of dimension ($6\text{ mm} \times 5\text{ mm} \times 0.5\text{ mm}$), gap between two slits 8 mm and the detectors inside the shielding boxes as shown in Fig. 5 – minimizes scattering effects.

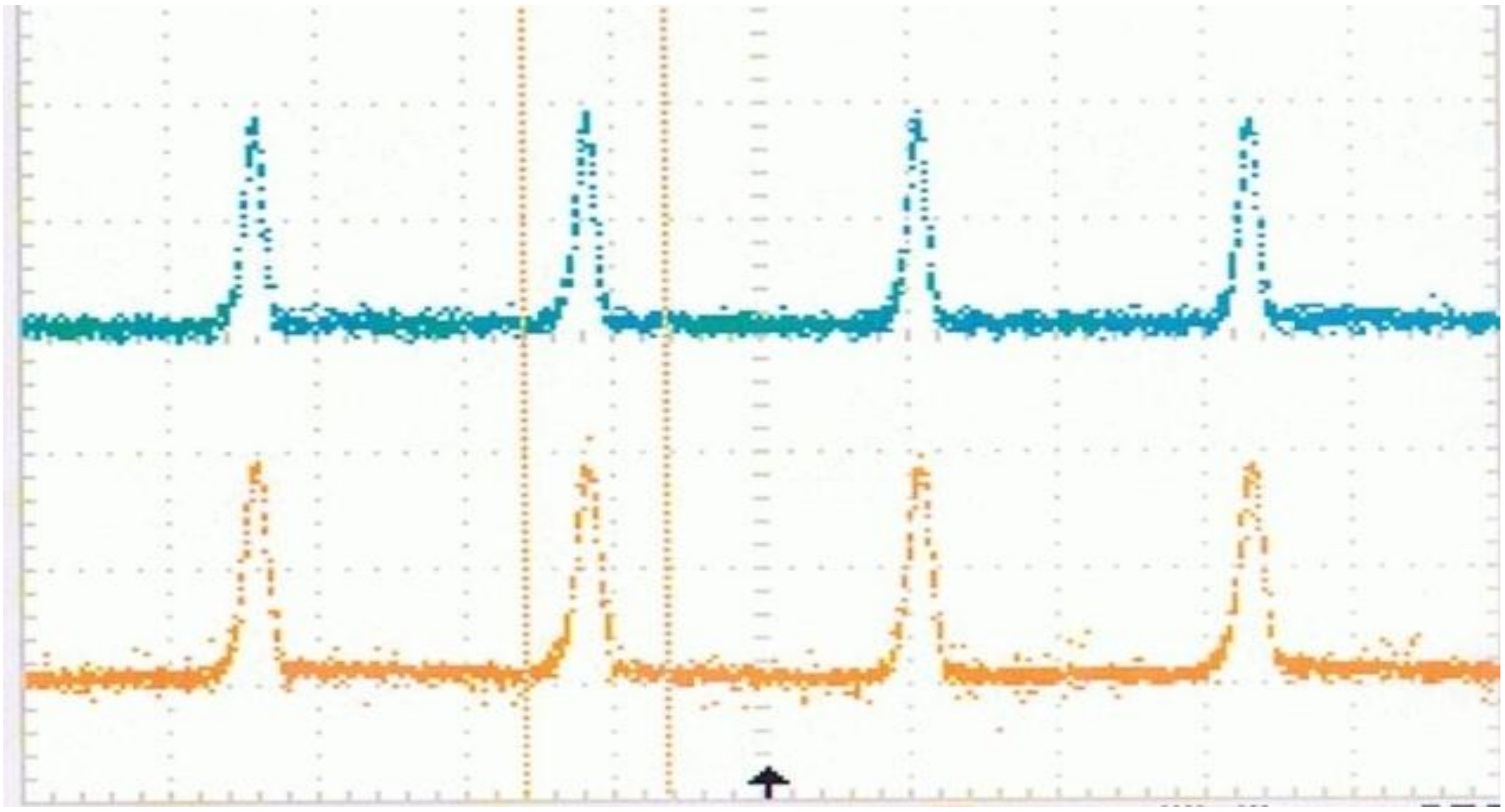

Fig. 7 Oscilloscope's [TDS2000B Digital Storage Oscilloscope] signals due to the responses of the two detectors (D1 and D2) in Fig 6: both Holed Disks (HD) open and close at the same time. Both channels have same setting (10 mV/div along vertical and 2.50 ms/div along horizontal).

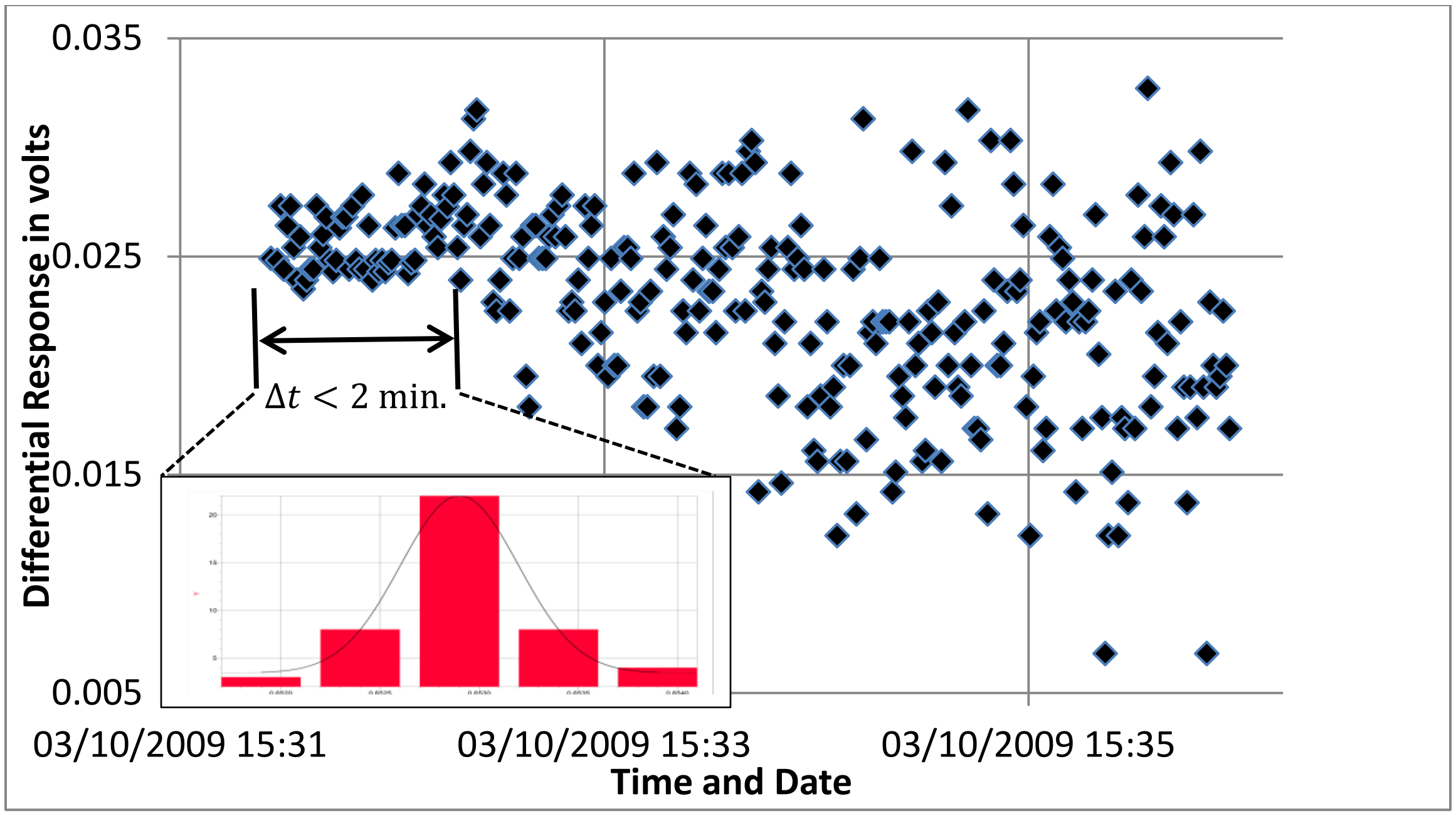

Fig. 8 Differential responses recorded in the computer according to our approach as described in Fig. 6. Laser was locked according [20] and Motor speed was 3200 RPM. The most reliable data can be found when run the time is $\Delta t < 2$ minutes. Histogram of this best data is shown.

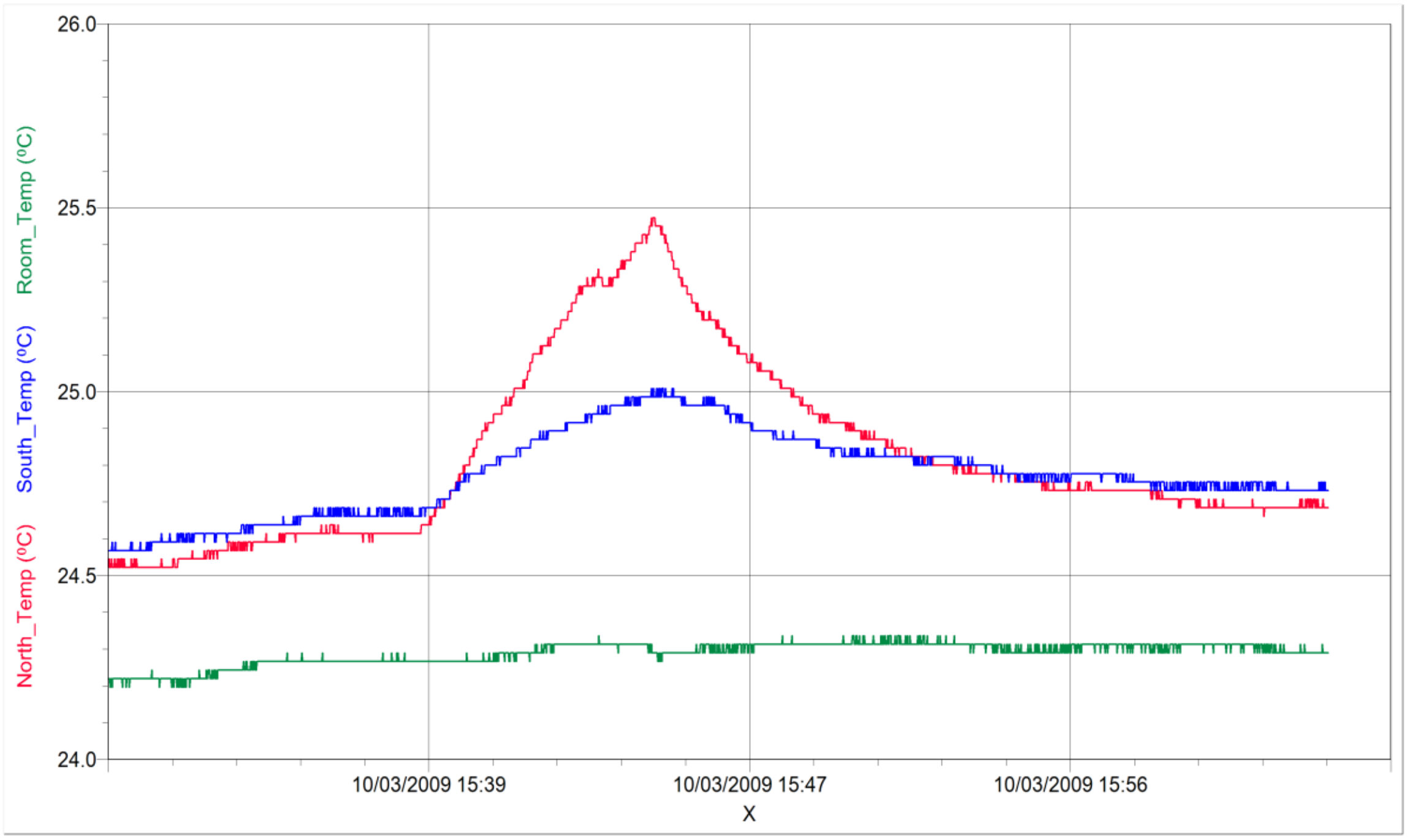

Fig. 9 Variation of temperatures during the run presented in Fig. 8. Top signal presents the temperature change near to the North detector (near to the motor), middle signal presents the temperature change near to the South detector and bottom signal represents the room temperature. This explains the deviations in differential responses presented in Fig. 8 after 2 minutes.

# 4. RESULTS AND DISCUSSION

## 4.1 The early setup and the direction for further understanding:

Marinov devoted himself trying to establish the absoluteness of space-time [26 - 32] and also to measure the absolute velocity of the solar system by means of a coupled mirrors experiment [9, 33] and a coupled shutters experiment [9 – 11, 33, 34]. The latter references presented controversial results compared with other established results [2, 35]. However, Marinov [9 – 11] emphasized the need to repeat the experiment. Reports in [2, 12, 36 - 38] also suggested repeating this experiment in a sophisticated laboratory. Despite these requests, discussions and criticisms, there has been no repetition so far. In our early setup, we followed as closely as possible Marinov's methods [9-11] and were able to collect differential responses and results exhibiting variations similar to Marnov [9 – 11]. However, our analysis and experimentation indicates that it is challenging to use this early setup to understand whether variations are due to the claim by Marinov that the speed of light is anisotropic or whether they are due to diurnal disturbances. It is unclear which methods Marinov utilized for controlling variations in photo detectors response caused by spatial and temporal variations of temperature, pressure and humidity, and scattering and, consequently it is difficult to duplicate the original setup precisely.

## 4.2 The improved setup and the preliminary studies

In order to get a deeper understanding of the experiment and its outcome and to ensure a valid result, we made improvements in the setup, experimental procedures and theoretical interpretation as described in the previous sections. This improved setup is able to collect data not only for the differential responses but also for the individual signals (North and South propagation for our experiment) as well as the control responses. The outcome of the improved setup will be published in near future. But we present a brief description of the preliminary studies in this subsection which can help one to identify the errors and analysis procedure to correct the errors.

In order to get an idea about the relationship among all responses in our improved setup, we present an example based on data for the responses which were collected before and after locking the laser. Fig. 10a presents the relationship among the responses before locking the laser as a diagnostic to indicate if the laser is stable. The observations of the diurnal variations in the speed of light (if any) due to the rotation of the Earth and identification of that due to the disturbances are prime objective of this work. In order to understand the variation due to any diurnal disturbances in laser stability – following [19] and variations exemplified by Fig. 10a – let us consider, for example, approximate equations where there is no phase difference as follows:

$$\mathcal{R}_N = \mathcal{A}_1 \sin\left(\frac{2\pi}{T} t + \alpha\right) \tag{23}$$

$$\mathcal{R}_S = \mathcal{A}_2 \sin\left(\frac{2\pi}{T} t + \alpha\right) \tag{24}$$

$$\mathcal{R}_{Diff} = [\mathcal{A}_1 - \mathcal{A}_2] \sin\left(\frac{2\pi}{T} t + \alpha\right) \tag{25}$$

$$\mathcal{C} = \mathcal{B} \sin\left(\frac{2\pi}{T} t + \alpha\right) \tag{26}$$

where $\mathcal{R}_N$ =North response, $\mathcal{R}_S$ = South response, $\mathcal{R}_{Diff}$ = Differential response, $\mathcal{C}$ = the control, $\mathcal{A}_1$ = Amplitude for North response, $\mathcal{A}_2$ = Amplitude for South response and $\mathcal{B}$ =Amplitude for the control.

Ideally, if we consider $\mathcal{A}_1 = \mathcal{A}_2$ then using equation (25) we get,

$\mathcal{R}_{Diff} = [\mathcal{A}_1 - \mathcal{A}_2] \sin\left(\frac{2\pi}{T_k} t\right) = 0.$

Here, there will be no observed variations in the differential responses.

However, in practice, it is challenging to make $\mathcal{A}_1 = \mathcal{A}_2$. Therefore, we observe sinusoidal diurnal variation of the differential signal due to the laser instability for diurnal disturbances where $T_k = T = 24$ hrs. Also, the identical variation can be observed in the individual responses as well as the control response which we are able to detect in our improved setup.

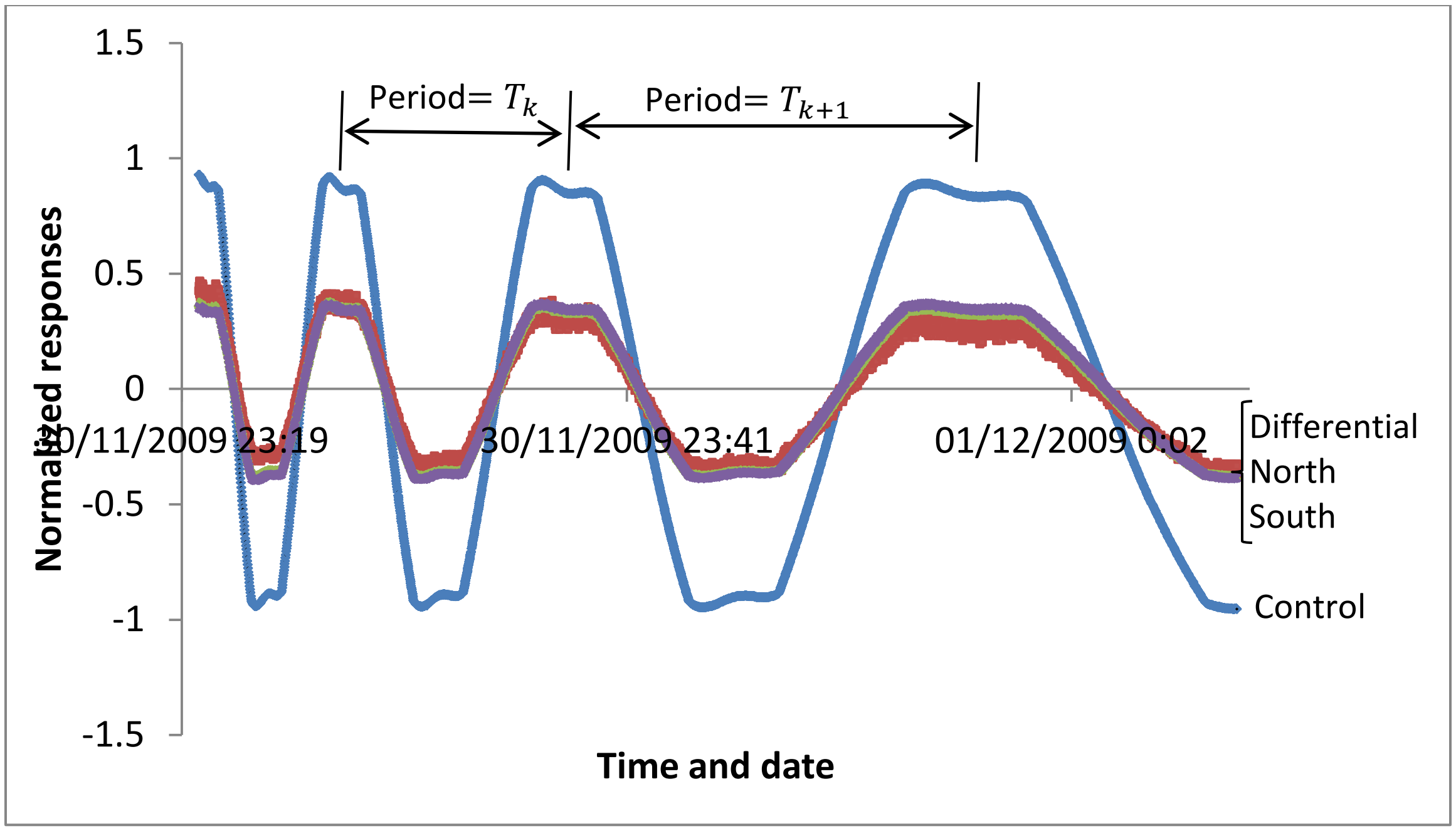

Fig. 10a The relationship among responses of the North detector, South detector, Differential (North – South) and the Control before locking the laser (i.e., when laser is unstable) and the motor speed =0. The flat potions at the maxima and minima are due to cutoff based on the setting to collect the data into the computer. The periods of the cycles are changing with the relation $T_k < T_{k+1}$ during warm-up time just after turning on the laser.

In order to correct the variation due to the laser instability, we can divide the responses derived in equations (23), (24) and (25) by the control equation (26) and we can get the corrected responses as follows:

$$\mathcal{R}_N/\mathcal{C} = \mathcal{A}_1/\mathcal{B} \quad (27)$$

$$\mathcal{R}_S/\mathcal{C} = \mathcal{A}_2/\mathcal{B} \quad (28)$$

$$\mathcal{R}_{Diff}/\mathcal{C} = (\mathcal{A}_1 - \mathcal{A}_2)/\mathcal{B} \quad (29)$$

Equations (27), (28) and (29) represent the corrected responses and also what we observe which is shown in Fig. 10b when the laser is locked and stable.

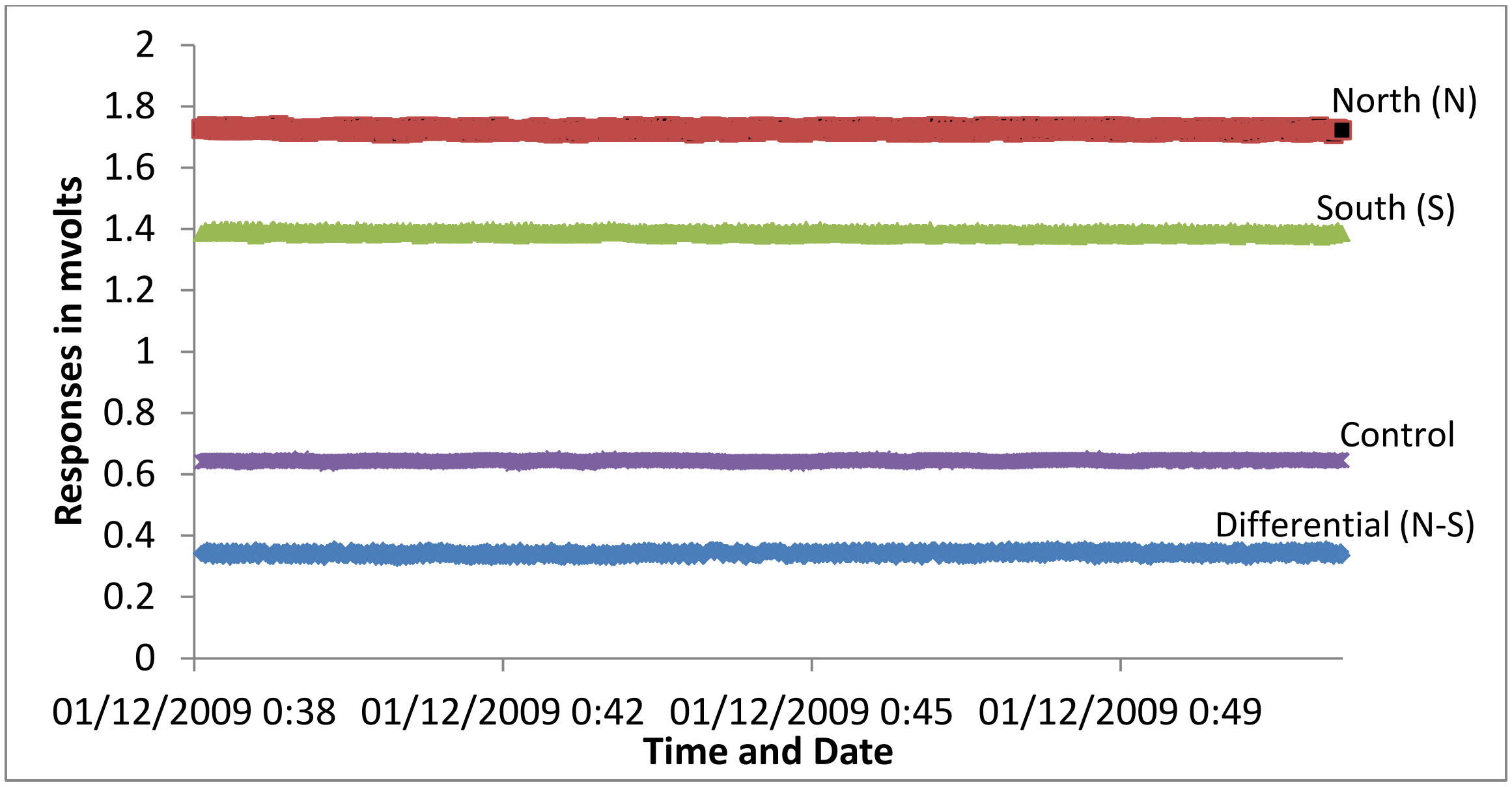

Fig. 10b The responses after locking the laser with the speed of the motor = 0.

If Marinov's claim [9 – 11] is correct (that, the speed of light is anisotropic), then what kind of variation might we expect in the outcome of our experiment? In order to determine the answer to this question, we present the predicted results in Fig. 11 which are derived following [2, 25] and using equations (17), (18) and (19) assuming a Galilean transformation.

In order to understand further, let us consider, for example, similar equations (23 – 25) but with a phase difference $\pi$ between two opposite responses as follows:

$$\mathcal{R}_N = \mathcal{A}_1 \sin\left(\frac{2\pi}{T} t + \alpha\right) \tag{30}$$

$$\mathcal{R}_S = \mathcal{A}_2 \sin\left(\frac{2\pi}{T} t + \pi + \alpha\right) = -\mathcal{A}_2 \sin\left(\frac{2\pi}{T_k} t + \alpha\right) \tag{31}$$

$$\mathcal{R}_{Diff} = [\mathcal{A}_1 + \mathcal{A}_2] \sin\left(\frac{2\pi}{T} t + \alpha\right) \tag{32}$$

Comparing equations (23 - 25) with equations (30 - 32), and also Fig. 11, we can easily identify the difference between the hypothetical variation and the variation due to the diurnal disturbances. The control (equation (26)) can be used to correct any variations due to diurnal disturbances.

Following [2, 25] and using equations (17), (18) and (19) we can also put limits on the velocity of the laboratory at any place on Earth based on the results of any experiment. For example, if we consider the data presented in the Fig. 11 and we assume that this represents the response due to the anisotropy of the speed of light in our experiment at York University then the limit of the velocity of the laboratory in different time on May 15, 2009 can be presented as in Fig. 11.

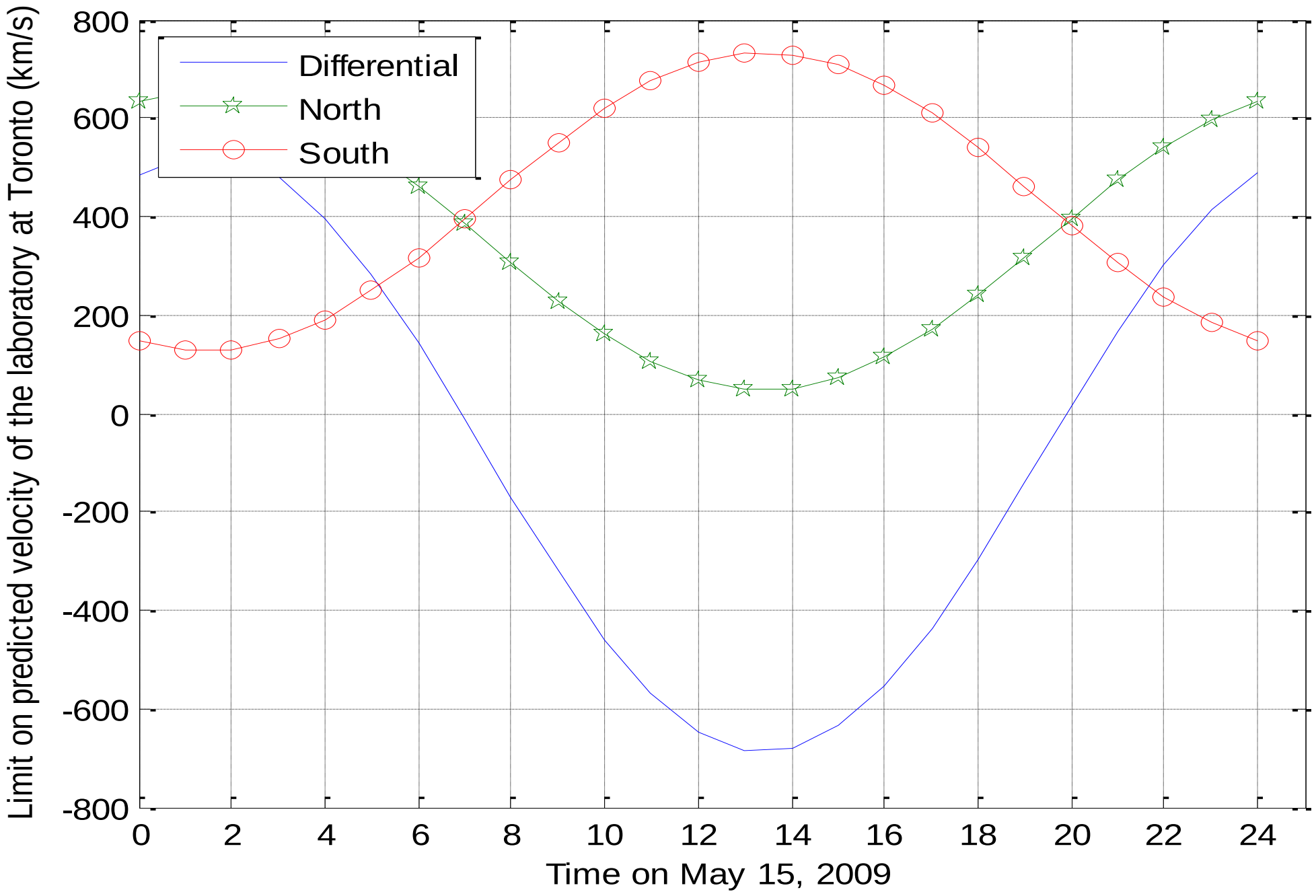

Figure 11: The limit of the velocity of the laboratory at Toronto with respect to the Cosmic Microwave Background (CMB) for 24-hours following [2, 25].

We will present initial data collected from our setup where the propagation of light was along the North-South directions. This direction was chosen following Marinov [9 – 11]. However, equations (17 – 19) indicate that the present experiment is insensitive to rotation of earth in the direction of the spin-axis. Thus if the earth were moving in this direction we would not be able to detect it.

## 5. CONCLUSION

The controversy about the results on the limits of the isotropy of the one-way speed of light from NASA-experiments [2, 7] and the regularity in the variations of the reported results of the isotropy of the one-way speed of light in different time periods of the GRAAL facility of the European Synchrotron Radiation Facility (ESRF) in Grenoble [39 – 43] remain unclear and require further investigation by one-way experiments. These previous experiments have yet to be repeatable by different laboratories as is possible with Michelson-Morley type two-way experiments. In order to look for a one-way experiment to test the isotropy of the speed of light that is repeatable and also, to ensure the validity of results, we are responding to proposals by several authors to repeat the Fizeau type experiment [2, 12, 36 - 38].

We have presented an improved version of the simple one-way speed of light experiment. The beauty of this experiment is its simplicity. According to our theoretical interpretation and experimental design, our approach is unique compared to other traditional approaches. In addition to measuring the variation on the one-way speed of light, we can determine the limits on the accuracy of our results. We express our initial calibration results in terms of physically measurable quantities which have been collected for a 24-hour period and compared each other under identical conditions. One-way isotropy measurement results from this experiment are forthcoming.

## Acknowledgments

The authors are thankful to Dr. Eamonn McKernan, Thoth Technology Inc., Mr. Nick Balaskas, Department of Physics and Astronomy, York University, Toronto, Canada. This work was supported by National Research Council of Canada, York University and Thoth Technology Inc. which are greatly acknowledged.